\documentclass[
reprint,
groupedaddress,
bibnotes,amsmath,amssymb,aps,
floatfix,]{revtex4-1}

\usepackage{array}
\usepackage{tabularx}
\usepackage{graphicx}
\usepackage{dcolumn}
\usepackage{bm}

\usepackage[utf8]{inputenc}
\usepackage{color}
\usepackage{graphicx}
\usepackage{amsmath}
\usepackage{natbib}

\begin{document}

\title{Probability distribution functions of sub- and super-diffusive systems}
\author{Fabio Cecconi}
\affiliation{Istituto dei Sistemi Complessi-CNR, Via dei Taurini 19 and 
INFN Unit\`a di Roma1, I-00185 Rome, Italy}
\author{Giulio Costantini}
\affiliation{Istituto dei Sistemi Complessi-CNR, 
Sapienza, Piazzale Aldo Moro 2, I-00185 Rome, Italy}
\author{Alessandro Taloni} 
\affiliation{Istituto dei Sistemi Complessi-CNR,Via dei Taurini 19, I-00185 Rome, Italy}
\author{Angelo Vulpiani}
\affiliation{Dipartimento di Fisica, Universit\`a Sapienza, Piazzale 
Aldo Moro 2, I-00185 Rome, Italy}

\date{April 12th 2021}

\begin{abstract}
We study the anomalous transport in systems of random walks (RW's) 
on comb-like lattices with fractal sidebranches, 
showing subdiffusion, and in a system of Brownian particles driven by a random shear along 
the $x$-direction, showing a superdiffusive behavior.
In particular, we discuss whether scaling and universality are present or not in the shapes of the particle distribution along the preferential 
transport direction ($x$-axis).
\end{abstract}

\maketitle

\section{Introduction}
Standard diffusion is a Gaussian behavior characterized by a
linear time growth of the mean square displacement (MSD) from
the initial condition
$$
\langle [x(t) - x(0)]^2\rangle \sim t\;.
$$
However certain processes in nature as well as in finance and even in sociology, 
are not Gaussian showing a nonlinear time growth of the 
MSD: 
\begin{equation}
\langle [x(t) - x(0)]^2\rangle \sim t^{2\nu}\;.
\label{eq:anomalous}
\end{equation}
If $\nu <1/2$ the process $x(t)$, by definition, undergoes a sub-diffusive behavior while, if $\nu>1/2$, $x(t)$ is said to be super-diffusive.

There is not a unique origin of the anomalous diffusion; it can be 
due to many different causes which are specific to the process under investigation. 

From the mathematical side, anomalous diffusion is a strict consequence of the breakdown of the Central Limit Theorem hypothesis, which can be ascribed either to the broad distributions of independent elementary steps, e.g. L\'evy flights \cite{levyflights}, or to the emergence of long-range, spatial or temporal, correlations among such steps.

In transport processes, the main subject of this paper, spatial correlations may arise from strong geometrical confinement and inhomogeneity due to the presence of disorder, obstacles, compartments and trapping sites, that can be found in amorphous \cite{street87,Amorph} and porous \cite{Berkowitz2000,Koch88} materials. 
Similarly, the crowding of cellular cytoplasm \cite{Cytoplasm,Cell_Transp} and organic tissues \cite{kopf1996anomalous,Tortuosity2} can make the motion of molecules and water in biological environments strongly correlated.
Correlations among consecutive displacements can also emerge dynamically \cite{gabrielli07}, as in the case of chaotic dynamics~\cite{Geisel84,klages_AnTraBook,vollmer21}, or because the particles are driven by the motions of the underlying medium, e.g., diffusion in turbulent fluid. Richardson, for instance, proved that in a turbulent fluid, spatial dispersion of particles follows a super-diffusive behavior \cite{Richardson26,boffetta_sokolov}. 

Similar situations occur when the random walks (RWs) are restricted on peculiar topological structures, such as fractals and networks
\cite{havlin_AdvPhys,weiss_comb86,Weiss_Shlomo87},  
where the geometrical constraints do not allow a fast memory loss among a series of consecutive displacements.

Finally, it is important to remark that often, in real physical and 
biological cases \cite{suppressDiff}, the anomalous diffusion can manifest 
itself as a transient phenomenon that, although long-lived, soon or later 
either turns into a standard one, or even stops due to the finiteness of the environments.

The goal of this paper is to discuss the general scaling properties of the probability distribution function (PDF) of anomalous diffusion at large times.
In analogy with standard diffusion which undergoes the natural scaling
\begin{equation*}
P(x,t) \propto  t^{-1/2}\exp[-c (x/\sqrt{t})^2]\;,
\label{eq:gauss}
\end{equation*}
we wonder to what extent the conjecture     
\begin{equation}
P(x,t) \sim t^{-\nu}F_{\nu}(|x|/t^{\nu})
\label{eq:PDFscale}
\end{equation}
can hold for sub- and super-diffusive processes. 
Obviously, the consequent issue is to determine the shape of 
$F_{\nu}(z)$ and understanding its degree of universality. 

At first, we have to bear in mind that the universality of this scaling is certainly broken by the existence of diffusion phenomena classified as ``strongly anomalous'' \cite{Castiglione99,andersen20simple,vollmer21}. 
Indeed, they exhibit a multiscaling in the spectrum of moments, 
\begin{equation}
    \langle [x(t) - x(0)]^m \rangle \sim t^{m\nu(m)}\,,
    \label{eq:strong_anomal}
\end{equation}
with not constant $\nu(m)$, which is in contrast with Eq.\eqref{eq:PDFscale} because a single exponent is not sufficient to characterize the statistical properties of such processes. 
The conflict is evident by considering that Eq.\eqref{eq:PDFscale} prescribes the following scaling for the moments,  
\begin{equation}
M_m(t) = \langle |x(t)-x(0)|^m \rangle  \sim t^{m\,\nu}.
\label{eq:Mom_vs_t}
\end{equation}
In the following we will not address strongly anomalous phenomena.

Roughly speaking, there are two different scenarios that can be traced back 
to the property \eqref{eq:PDFscale}: one associated with time-homogeneous 
diffusion processes, for which the increment $x(t+h) - x(t)$ is independent of $x(t)$, 
and the other associated with processes not fulfilling the above property. The 
latter are said non-homogeneous processes and are especially observed in turbulent diffusion.

\begin{figure}
\includegraphics[width=0.85\columnwidth]{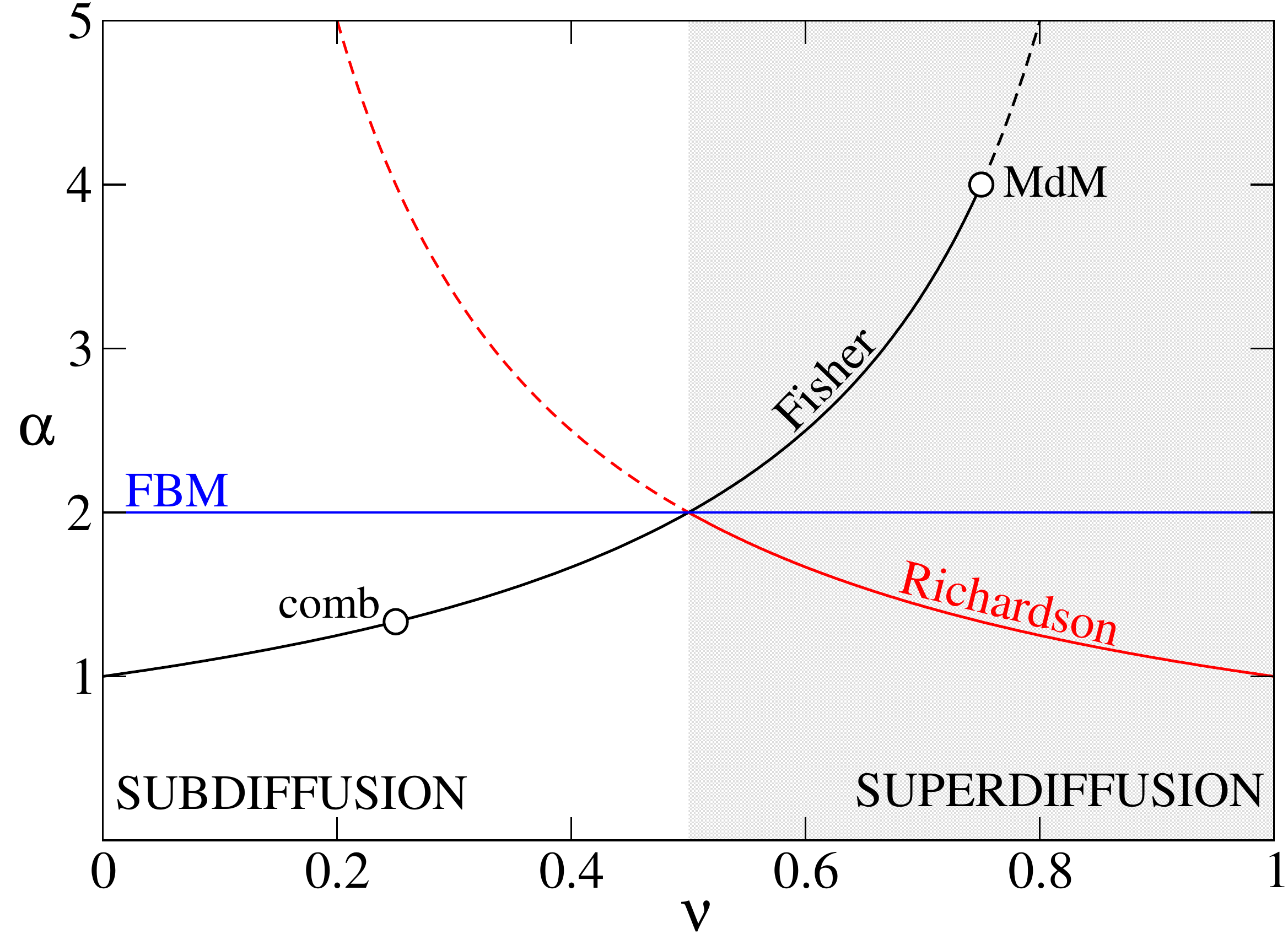}
\caption{Dependence of the parameter $\alpha$ of the stretched exponential 
as a function of the anomalous exponent $\nu$ Eq.\eqref{eq:branches}, 
defining the Fisher's and Richardson's branches. 
The horizontal line indicates the value $\alpha=2$ characterizing the fractional Brownian motion (FBM) and the intersection of all the lines 
($\nu=1/2,\alpha=2)$ identifies the Gaussian point. 
The point $(\nu=1/4,\alpha=4/3)$ corresponds to the simple comb model, and 
$(\nu=3/4,\alpha=4)$ corresponds to the Matheron de-Marsily (MdM) 
shear model \cite{mdem_model}. 
\label{fig:scenario}
}
\end{figure}

In the literature, some guesses can be found as to the functional
form of $F_{\nu}(z)$ depending on the anomalous exponent $\nu$. 
It is worth mentioning two classes of $F_{\nu}(z)$, the one suggested for the time-homogeneous processes (Fisher) and the one valid for turbulent diffusion 
(Richardson) both characterized by a stretched exponential 
\begin{equation}
F_{\nu}(z) \sim \exp(-a_{\nu}|z|^{\alpha}),
\label{eq:Stretched_exp}
\end{equation}
with exponents
\begin{equation}
\alpha =
\begin{cases}
\dfrac{1}{1-\nu} &  \mathrm{Fisher~branch} \\
                 &                   \\
\dfrac{1}{\nu}   &  \mathrm{Richardson~branch}.
\end{cases}
\label{eq:branches}
\end{equation}
Figure \ref{fig:scenario} sketches the expected validity range 
of the Fisher and Richardson $F_{\nu}(z)$ as a function of the anomalous exponents, in both sub- and super-diffusive regimes. 
It is interesting to note that for $\nu=1/2$, the two stretched 
exponentials become of Gaussian type, as $\alpha = 2$.
We refer to the $F_{\nu}(z)$ with $\alpha=1/(1-\nu)$ as the Fisher-like distribution, since Fisher derived it in the context of self-avoiding polymers (SAP) \cite{FisherSAP}. 
Whereas, the case of $F_{\nu}(z)$ with $\alpha=1/\nu$, which can be referred to as Richardson's function because it was derived by Richardson in his studies on turbulent diffusion \cite{Richardson26,boffetta_sokolov}, will not be addressed in this work.

Before discussing the numerical analysis we carried out for proving the Fisher's scenario, we present two arguments supporting the validity of   
\begin{equation}
\alpha = \dfrac{1}{1-\nu}.
\label{eq:fisheralpha}
\end{equation}
One is based on simple probability considerations \cite{redner92powlaw-shear}, 
and the other on the large-deviation theory (LDT) which has been already and successfully applied to describe the continuous time RW (CTRW) \cite{Sokolov_ctrwLDT,Barkai_ctrwLDT}.

The first heuristic argument starts from the observation that the main 
contribution to the tails arises from RWs that are very persistent along the 
$x$-direction, \emph{i.e.} $x_{max} \sim t$. 
Therefore, the probability of such persistent walks can be approximated as 
$F_{\nu}(z_{\mathrm{max}}) \sim \exp(-a_{\nu}|z_{\mathrm{max}}|^{\alpha})$, where delta time integration $z_{\mathrm{max}}=x_{max}/t^\nu \sim t^{1-\nu}$.  
On the other side, the probability of such paths is also 
$P\{z_{\mathrm{max}}\} \sim \exp(-p t)$, as they undertake  $n=t/\Delta t$ 
independent steps in the same direction, $p$ being a constant and $\Delta t$ the time step.
Equating $F_{\nu}(z_{\mathrm{max}})=P\{z_{\mathrm{max}}\}$ gives the Fisher's branch scaling prediction. 

The second heuristic argument is based on LDT, which is a natural probability 
framework to determine how large fluctuations from the mean of a random process characterize the decay of far tails of its PDF \cite{touchette}.
LDT is the proper mathematical tool if we are interested in verifying whether the far tails of the PDF \eqref{eq:PDFscale} are consistent with Fisher's tails.
LDT assumes that, at large time, the probability decays as   
\begin{equation}
P(x,t) \propto \exp[-t {\cal C}(x/t)]\,,
\label{eq:LDT}
\end{equation}
where ${\cal C}(\cdots)$ is the Cramers' (or rate) function.  
We recall that if a process $\mu_n$ is the average over $n$ 
independent random variables, $\mu_n = (\xi_1 +\cdots + \xi_n)/n$,   
its Cramers' function is, by definition, the limit  
${\cal C}(\mu_n) = -\lim_{n\to \infty} 1/n \ln P(\mu_n,n)$, implying that
asymptotically for large $n$
$$
P(\mu_n,n) \propto  \exp[-n {\cal C}(\mu_n)]. 
$$
This expression can be equivalently rewritten in terms of the sum $S = \xi_1 +\cdots + \xi_n = n\,\mu_n$ and the total time $t=n\Delta t$,
$$
P(S,t) \propto  \exp[-n {\cal C}(S/n)]=\exp[-t {\cal C}(S/t)]. 
$$
which is exactly Eq.\eqref{eq:LDT}.   
A comparison between Eq.\eqref{eq:LDT} and Eq.\eqref{eq:Stretched_exp}
implies that
$ t {\cal C}(x/t) = a_{\nu}|x/t^{\nu}|^{\alpha}$, this equality can 
be rearranged to ${\cal C}(x/t) = a_{\nu}|x/t^{(\nu+1/\alpha)}|^{\alpha}$.
Now the only consistent possibility is that ${\cal C}(x/t) = a_{\nu}|x/t|^{\alpha}$ 
which can only hold true if Eq.\eqref{eq:fisheralpha} is verified.

Of course, although the above reasoning does not constitute a proof, yet we can conclude that Fisher's PDF is the unique PDF which is consistent with the scaling \eqref{eq:PDFscale} and the LDT as well.
   
\begin{figure}[ht!]
\centering
\includegraphics[width=0.6\linewidth]{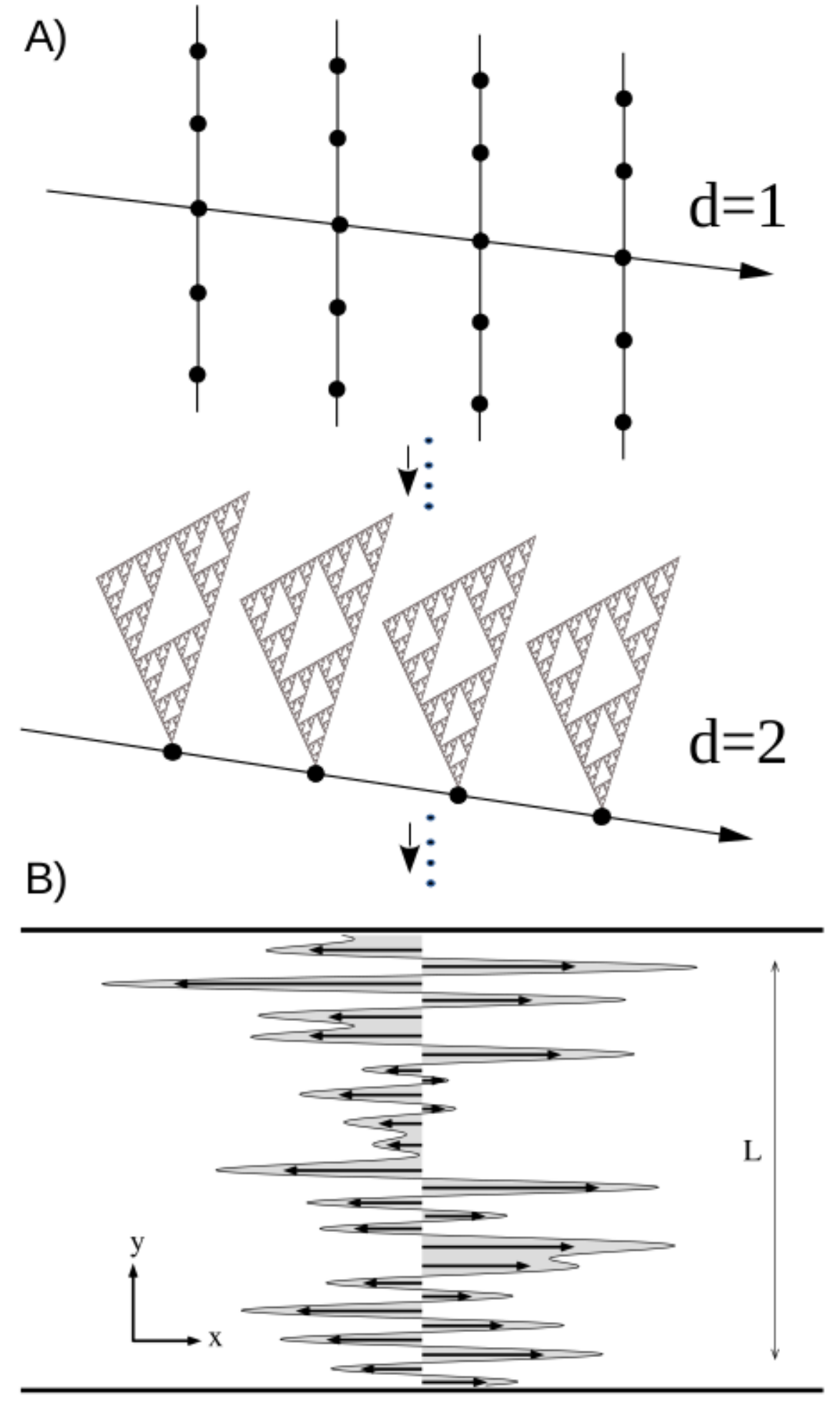}
\caption{(A) Cartoon of fractal structures studied in this work: 
simple comb lattice and comb of Sierpinski gaskets. 
	(B) Sketch of the two-dimensional (2D) random velocity
field, $\mathbf{U}=[U(y),0]$, parallel to the $x$-axis and depending only 
on the vertical coordinate $y$.}
\label{fig:cartoon}
\end{figure}

In this paper we will focus only on the validity of the Fisher's scenario in the subdiffusive and superdiffusive regimes, by using two models.\\
(1) The random walks (RWs) on a class of comb-like structures  consisting of a main backbone decorated by an array of fractal sidebranches \cite{Burioni05}, as in Fig.\ref{fig:cartoon}(A).
In this case, the subdiffusion is observed along the backbone. \\
(2) The Lagrangian dynamics of particles in a channel geometry under the combined effects of a random velocity shear and molecular diffusivity, Fig.\ref{fig:cartoon}(B) \cite{dreizin1973anomalous,mdem_model}. 
In this case, the superdiffusion occurs along the longitudinal transport direction ($x$-axis).
We will show that the scaling behavior \eqref{eq:PDFscale} is well satisfied by both the sub- and super-diffusive processes that we studied.
This is testified by the perfect collapse, upon re-scaling $x\to |x|/t^{\nu}$, of the simulated particle distribution (PDF) onto a master curve whose tails can be described by Eq.\eqref{eq:Stretched_exp} with exponent \eqref{eq:fisheralpha}. 
As a matter of fact we will demonstrate, numerically and analytically, that the Fisher relation is rigorously satisfied in the case of comb-like structures, even when the sidebranches are fractal structures of growing complexity (Sierpinski gaskets with increasing spectral dimension, Fig.\ref{fig:cartoon}(A). 

On the other side, the superdiffusive case of the random-shear model [Fig.\ref{fig:cartoon}(B)] presents an unexpected scenario where the Fisher scaling appears to be satisfied only for a restricted range of the anomalous exponent.

Moreover, the behavior of the PDF moments in agreement with  Eq.\eqref{eq:Mom_vs_t} provides further numerical support to the validity of the re-scaling, Eq.\eqref{eq:PDFscale}.

 As we will see in the following, the very numerical difficulty in observing the true nature of the tails stems from the necessity of two simultaneous asymptotic conditions, $t\gg 1$ and also $|x|/t^{\nu}\gg 1$. 

This paper is organized as follows. In Sec.\ref{sec:comb}
we study the anomalous behavior of RWs on comb-like structures with fractal sidebranches.
In Sec.\ref{sec:shear}, we analyze the numerical results on the  super-diffusion of Brownian particles under a random shear in a channel of width $L$.

Finally Sec.\ref{sec:concl} contains a discussion and conclusions.

\section{Comb-like systems
\label{sec:comb}}
To study the scaling of the PDF of anomalous sub-diffusion, 
we considered RW's on comb-like structures, namely, a central backbone decorated by either linear or more complex sidebranches (SB's) \cite{weiss_comb86}. 
In particular, we are interested in fractal structures, so we consider SB's constituted of Sierpinski-gaskets with increasing spectral dimension [see Fig.\ref{fig:cartoon}(A)]. 
Such branched topology is typical of percolation clusters at criticality, which can be viewed as finitely ramified fractals \cite{coniglio82,conigliostanley}.
Comb-like structures, moreover, are frequently used in condensed matter and biological frameworks to describe the geometry of branched polymers \cite{polycomb,polybranch}, amphiphilic molecules and engineered structures at the nano- and microscales, and anomalous propagation of chemical reaction fronts \cite{mancinelli03front}.
Moreover, in recent years a series of papers focused on the generalizations of the original comb model, as systems encompassing superdiffusion, due to the presence of inhomogeneous convection \cite{iomin2005negative} or L\'evy flights \cite{sandev2016levy}, heterogeneous and fractional diffusion on a fractal mesh \cite{sandev2015fractional,sandev2018heterogeneous,sandev2017anomalous}, and slow or ultraslow diffusion \cite{sandev2016comb}.
In general, comb models can be seen as a concrete representations of CTRW. A general account of these systems can be found in Ref.~\cite{Ben-Avraham}. 

The analysis of the diffusion along the backbone has been obtained by performing numerical simulations of $N = 3\times 10^6$ RWs over a Sierpinski comb-structure, with one gasket for every backbone site.
We consider RWs on several comb-structures differing in the fractal dimension of the SBs. Each SB is characterized by a primary element by means of which one can iteratively generate the fractal structure. Such a geometrical element is 
called $\delta$-simplex, which is a set of $\delta$ sites joined by edges. 
The Euclidean dimension $d$ of the space in which the gasket is embedded is related to the simplex by $d=\delta-1$ \cite{gasketRW}.
The spectral dimension $d_s$ of such structures is related to the Euclidean dimension by the relationship \cite{Alexander,gasketRW}
\begin{equation}
d_s = 2\,\dfrac{\ln (d+1)}{\ln(d+3)},
\label{eq:ds}
\end{equation}
while the fractal dimension $d_f$ depends on $d$ via the relationship
\begin{equation}
d_f = \dfrac{\ln(d+1)}{\ln(2)}.
\label{eq:df}
\end{equation}
The behavior of the mean-square displacement (MSD) of the RW over
comb-like structures is anomalous, Eq.\eqref{eq:anomalous}, 
with an exponent $\nu$ depending on the spectral dimension of the SB:
\begin{equation}
\nu = \dfrac{2-d_s}{4}.
\label{eq:fract}
\end{equation}
Of course, the unavoidable finiteness of the linear size $L$ of the SB's makes the anomalous diffusion a transient behavior occurring before the onset of a standard regime.
However, upon taking $L$ to be sufficiently long, with respect to the mean free path, the transient anomalous behavior can be made arbitrarily long-lived.

Equation \eqref{eq:fract} can be explained by a simple phenomenological argument \cite{Forte2013} which is based on the 
{\em homogenization time}, $t_*(L)$, meant as the typical timescale after which the diffusion along the backbone becomes standard 
\begin{equation}
\langle x_{\parallel}^2(t) \rangle \sim 
\begin{cases}
t^{2\nu}     & \mbox{if  $t\ll t_*(L)$} \\
D(L)\,t      & \mbox{if  $t\gg t_*(L)$} 
\end{cases}
\label{eq:arg2}
\end{equation}
where $D(L)$ is the effective diffusion coefficient depending on the scale $L$. For finite-size SB's indeed, the anomalous regime in the longitudinal diffusion is only transient, and soon or later, it will be replaced by the standard diffusion.
The homogenization time, $t_*(L)$, can be identified with the typical time taken by the walker to ``span'' a single SB of linear size $L$.
The scaling of $t_*(L)$ with $L$, can be easily extracted from the diffusion on fractal structures, $\langle x^2_{\perp}(t)\rangle \sim t^{2/d_w}$, where 
$d_w$ indicates the random walk dimension \cite{Ben-Avraham}. Therefore we expect that, $L^2 \sim [t_*(L)]^{2/d_w}$, implying the scaling 
$$
t_{*}(L) \sim L^{d_w}\;.
$$
Once such a scaling is known, we can apply a ``matching argument''
to derive the exponent $\nu$ in Eq.\eqref{eq:anomalous}, by requiring that 
the power-law and the linear behavior have to match at time $t \simeq t_*(L)$,
\begin{equation}
t_*(L)^{2\nu} \sim D(L)\:t_*(L)\;.
\label{eq:matching}
\end{equation}
We need to determine the scale dependence 
of the effective diffusion coefficient $D(L)$ for the RW along the backbone.
This scale dependence is 
\begin{equation}
D(L) \sim L^{-d_f},
\label{eq:fracD}
\end{equation}
$d_f$ being the fractal dimension of each SB. 
Indeed, when the diffusion along the backbone has reached the standard regime, 
it satisfies $\langle x_{\parallel}^2(t) \rangle = D(L) t$, where $D(L) = D_0 f(L)$ is the bare (microscopic) diffusion constant $D_0$ reduced by a factor $f(L)$, which is the probability for a walker to occupy a backbone site, since only the fraction $f(L)$ of RWs on the backbone actually contribute to the diffusion.  
We can safely assume that the dynamics in the homogenization regime equally visits all the sites on each SB, which become equiprobable.
Accordingly, $f(L)$ follows from a simple geometrical counting that is evident  
when referring to the simple one-dimensional comb with finite SB of length 
$L$, [top panel of Fig.\ref{fig:cartoon}(A)]. 
Clearly, each linear SB contains a number of sites $N_{\mathrm{SB}}(L)=L/\ell_0$, where $\ell_0$ is the lattice spacing that, without loss of generality   
can be set to $\ell_0=1$. 
Therefore, $f(L)$ amounts to the following simple counting: only one backbone site over $N_{\mathrm{SB}}(L) = L$ total SB sites , implying that $f(L) = 1/N_{\mathrm{SB}}(L) \sim L^{-1}$, in $d=1$.
The reasoning straightforwardly generalizes to fractal SB, [Fig.\ref{fig:cartoon}(A) bottom panel] by observing that, on a fractal, the number of sites scales 
as $N_{\mathrm{SB}}(L) \sim L^{d_f}$ \cite{havlin_AdvPhys}, thus we have again, $1$ backbone site over $L^{d_f}$ SB sites. 
The scaling \eqref{eq:fracD} indicates 
that $D(L)$ decays by enhancing the trapping power of the SBs, this occurs for two reasons: by increasing their linear size $L$, or by increasing their geometrical complexity, characterized by $d_f$.

Now, using $t_*(L) \sim L^{d_w}$, the matching Eq.\eqref{eq:matching} yields 
$2\nu d_w = -d_f + d_w$, from which we obtain Eq.\eqref{eq:fract}, 
providing that $d_w = 2d_f/d_s$, see Ref.\cite{havlin_AdvPhys}.
A similar reasoning has been generalized to compute anomalous exponents for RWs on fractal brushes, in Ref.\cite{brushes2017}.

\subsection{Comb and comb-Sierpinski structures 
\label{sec:comb_structures}}
\begin{figure}[ht!]
\centering
\includegraphics[width=0.85\columnwidth]{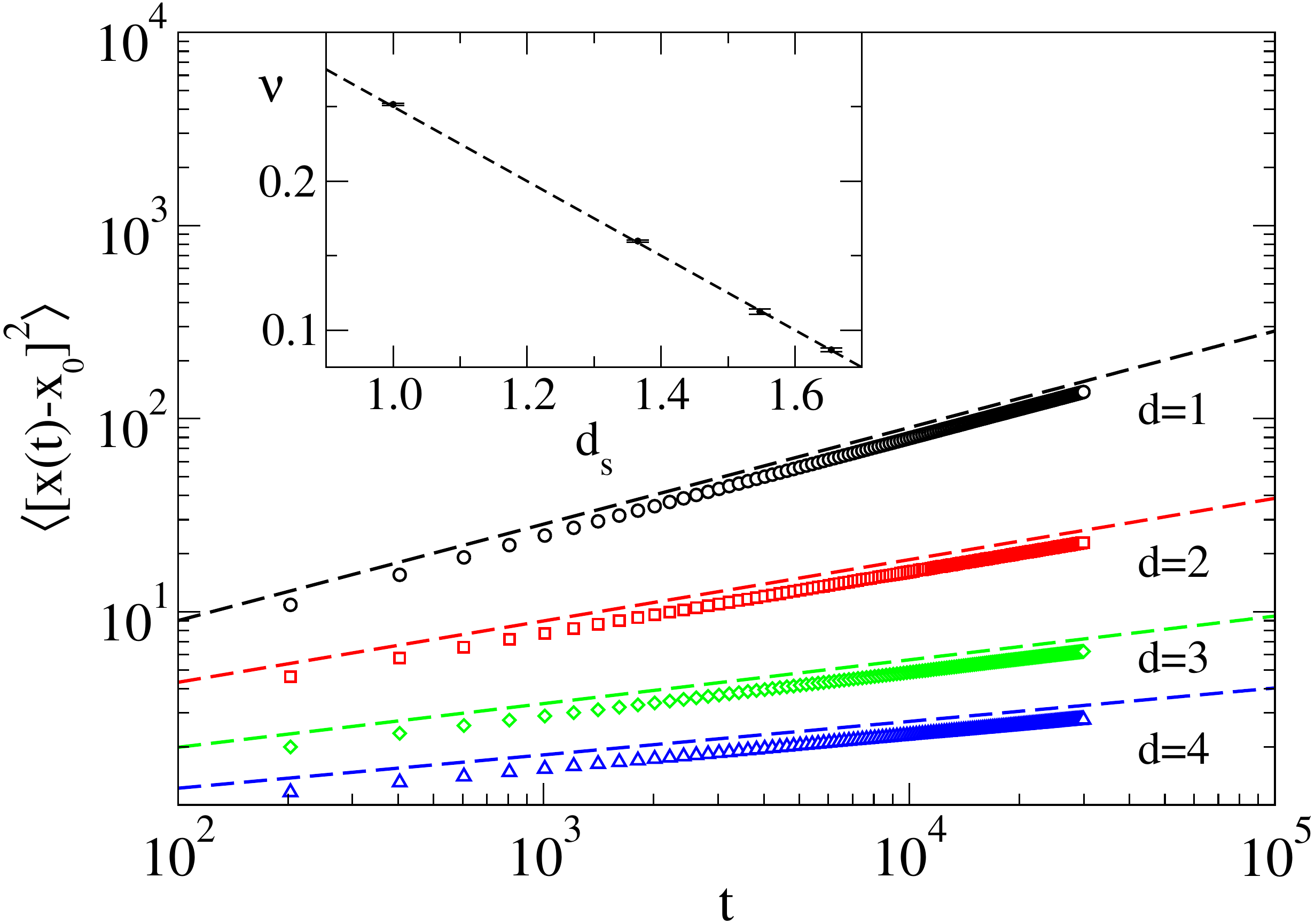}
\caption{Main plot: MSDs (symbols) vs. time for comb structures at different values of the spectral dimension: simple comb, $d_s=1$, (black symbols), $d_s=1.365$ (red symbols), $d_s=1.547$ (green symbols) and $d_s=1.654$ (blue symbols). 
	The dashed curves correspond to the MSD [Eq.\eqref{eq:anomalous}], with $\nu$ given by Eq.\eqref{eq:fract}. 
Inset: Plot of $\nu$ values (with errors), estimated from the slope of the MSD in the main panel, against the four considered values of $d_s$.  
The dashed curve represents Eq.\eqref{eq:fract}.
\label{fig:msd_vs_ds}
}
\end{figure}

\begin{figure}[ht!]
\centering
\includegraphics[width=0.85\columnwidth]{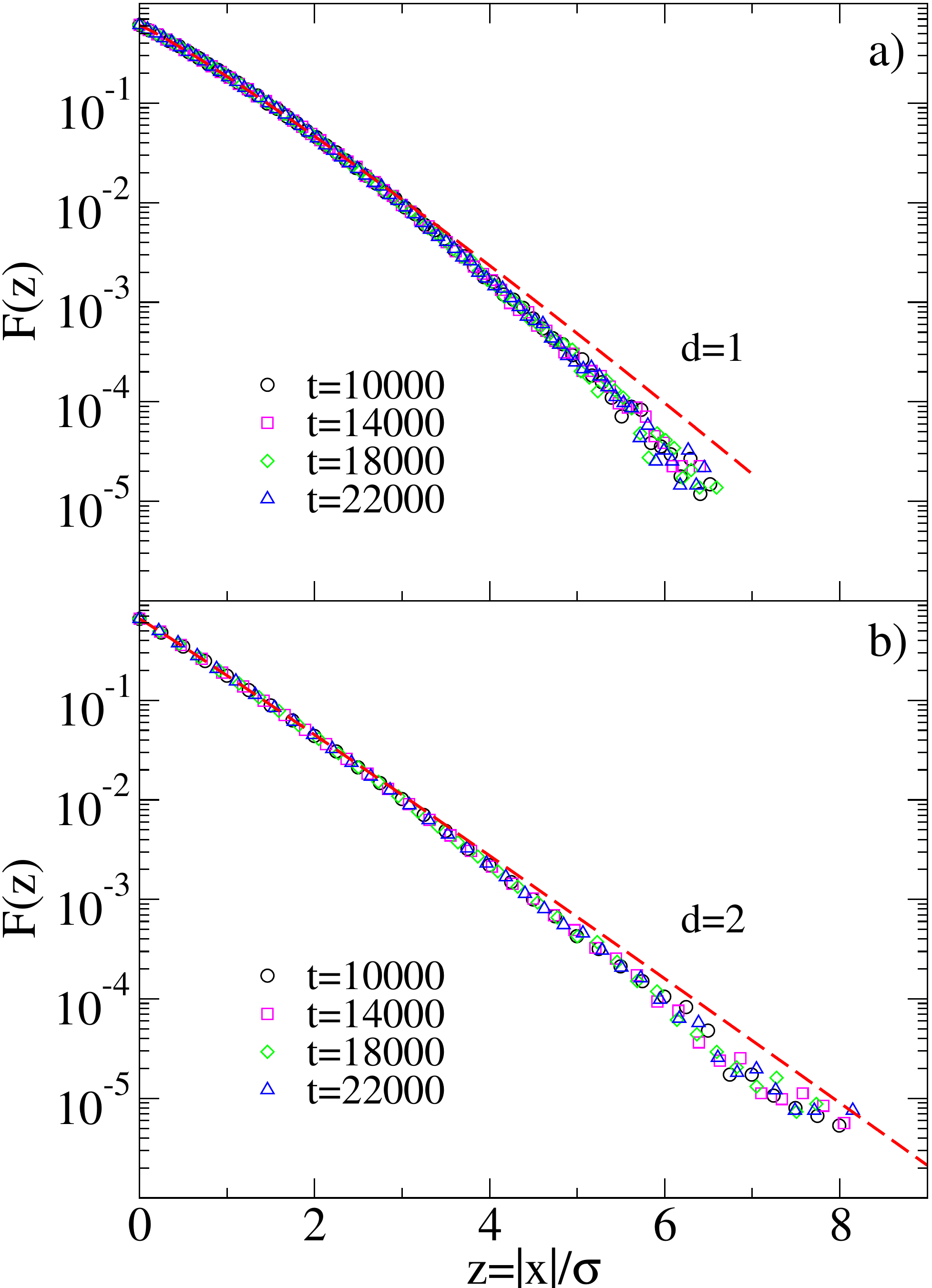}
\caption{Rescaled PDFs, $F(|x|/\sigma)$ of the comb-like structures for different values of time and different spectral dimensions: $d_s=1$ panel a) and $d_s=1.365$ panel b). The red dashed curve corresponds to the stretched exponential fit via Eq.\eqref{eq:Stretched_exp}.
}
\label{fig:pdfs_comb_rescaled}
\end{figure}

Because of the coupling between the SBs and the backbone dynamics, the transport along the backbone of a comb, see the top of Fig.\ref{fig:cartoon}(A), 
becomes anomalous with an exponent given by Eq.\eqref{eq:fract}. 
Figure~\ref{fig:msd_vs_ds} reports the power-law behavior of the MSD along with its prediction (dashed lines) for the structures corresponding to different $d_s$. 
The long-time PDFs of the RW dynamics on the backbone present
the collapse onto a $F_{\nu}(z)$ predicted by Eq.\eqref{eq:PDFscale}, 
upon rescaling, $z=|x|\sigma(t)$, with $\sigma(t) = \sqrt{\langle [x(t) - x(0)]^2}$, see Fig.\ref{fig:pdfs_comb_rescaled}.
To test whether the PDFs exhibit Fisher's tails, we performed a global fitting of all the curves using Eq.\eqref{eq:Stretched_exp} with $a_{\nu}$ and $\alpha$ as free parameters. 
This unconstrained or ``blind'' fitting procedure shows that the stretched-exponential form fits only the distribution bulk, i.e. small values of $z$. 
Moreover, the $\alpha$ values from the fitting deviate from the expected values, Eq.\eqref{eq:fisheralpha}, see Table \ref{tab:1}.  
This analysis suggests that without a proper guess, there is no chance 
to prove that the backbone diffusion in comb-like structures follows a Fisher-like distribution, at least on the tails.
\begin{table}[ht!]
{\begin{tabular}{cccccccccc}
&  & $d$ &  &  Fitted $\alpha$     & & $\alpha_F$ \\
\hline
&  & $1$ &  &  $1.1166 \pm 0.0023$ &  &  $1.3333$ & \\
&  & $2$ &  &  $1.0341 \pm 0.0010$ &  &  $1.1886$ & \\
&  & $3$ &  &  $1.0034 \pm 0.0049$ &  &  $1.1276$ &
\\
&  & $4$ &  &  $0.9776 \pm 0.0145$ &  &  $1.0946$ &
\end{tabular}
}
\caption{Values of the parameters of the best fitting of the simulated PDFs with Eq.\eqref{eq:Stretched_exp}.
}
\label{tab:1}
\end{table}
An analytical prediction of the PDF can be performed using the fractional Fokker-Planck equation (FFPE) which governs the diffusion on the comb backbone; see Ref.\cite{metzler2000random}. 

To start with, let us briefly recall the theory developed for the comb system in Refs.\cite{arkhincheev2007random,arkhincheev1991anomalous,arkhincheev2002diffusion}. 
According to the authors, the Smoluchowski equation for the particle distribution $P(x,y;t)$ on the comb (backbone plus SB's) is 
\begin{equation}
    \frac{\partial P(x,y;t)}{\partial t} = D_x\delta(y)\frac{\partial^2 P}{\partial x^2}+D_y\frac{\partial^2 P}{\partial y^2}.
\label{eq:FP_COMB}
\end{equation}
The presence of the $\delta(y)$ function ensures that the diffusion along the backbone takes place only at $y=0$. Two different diffusion coefficients are introduced: one along the comb teeth ($D_y$) and one along the backbone ($D_x$), with different physical dimensions. 
The equation for the particles diffusing along the backbone,
$$
P_{B}(x,t)=\int_{-\infty}^\infty dy\,P(x,y;t)\,,
$$ 
can be derived by manipulating Eq.\eqref{eq:FP_COMB} \cite{arkhincheev1991anomalous} 
defined as
\begin{equation}
    \frac{\partial P_B(x;t)}{\partial t} = K_{1/2}\frac{\partial^2 }{\partial x^2}\;{}_0D_t^{\frac{1}{2}}P_B(x;t),
\label{eq:FP_sol_x_total}
\end{equation}
where $K_{1/2}=D_x/(2\sqrt{D_y})$ and $ {}_0D_t^\alpha$ is the Riemann-Liouville fractional operator of order $\alpha$ \cite{podlubny1998fractional,metzler2000random}:

\begin{equation}
    _0D_t^{\alpha}\phi(t)=\frac{1}{\Gamma\left(1-\alpha\right)}\frac{d}{dt}\int_{0}^tdt'\frac{1}{\left(t-t'\right)^{\alpha}}\phi\left(t'\right)
\     \ 0<\alpha<1. \label{RL} 
\end{equation}

\begin{figure}[ht!]
\centering
\includegraphics[width=0.85\columnwidth]{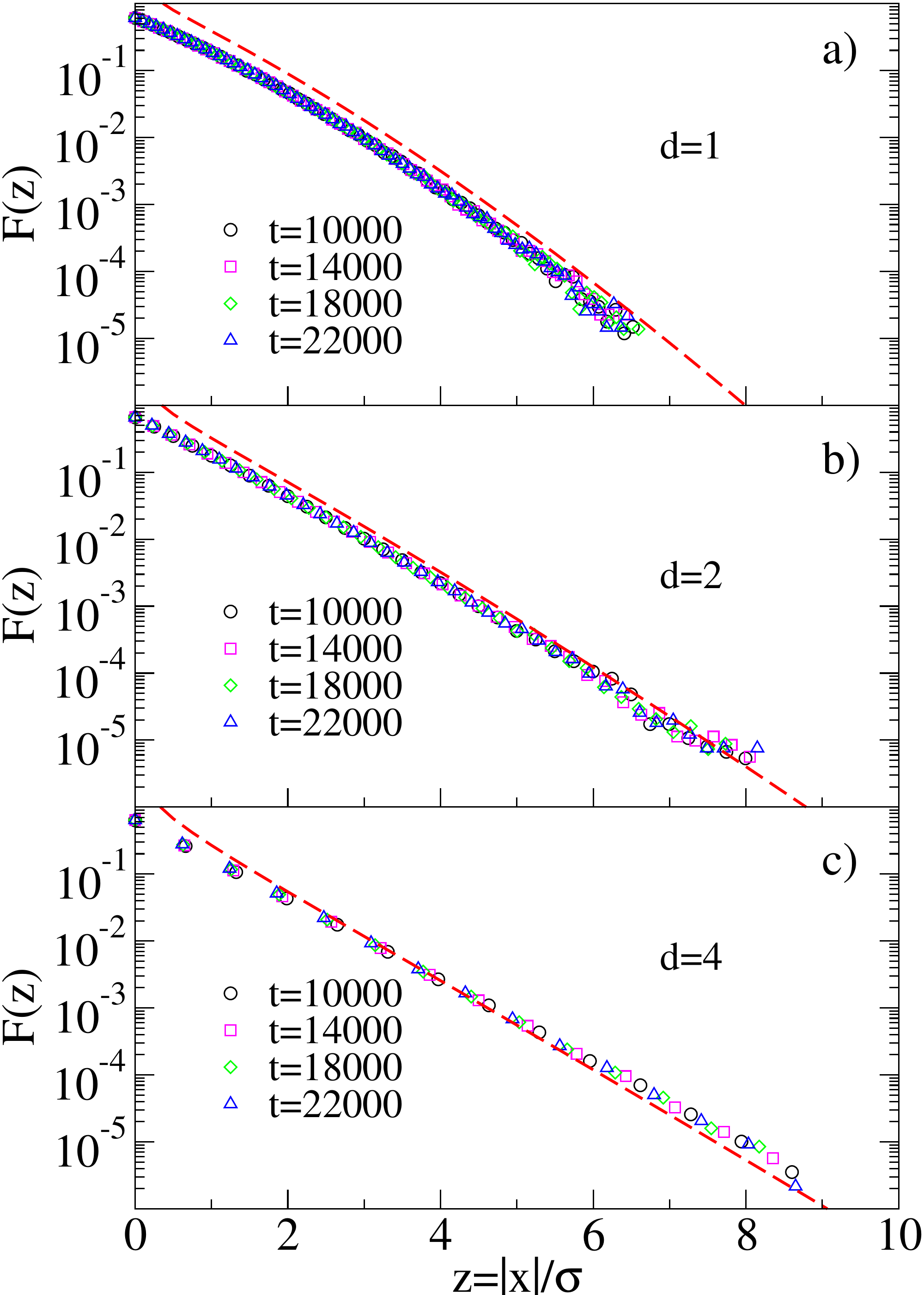}
	\caption{Rescaled PDF $F[|x|/\sigma(t)]$ of the comb-like structures for different values of time and different spectral dimensions: $d_s=1$ panel (a), $d_s=1.365$ panel (b) and $d_s=1.654$ panel (c). 
The red dashed curves are the plots of Eq.\eqref{eq:FP_sol_fisher} 
for panel (a) and Eq.\eqref{eq:FP_sol_fisher_COMB} for panels (b) and (c).
\label{fig:pdfs_Fisher_tails}. Notice that no fitting procedure is applied.
}
\end{figure}

As anticipated, Eq.\eqref{eq:FP_sol_x_total} is a type of FFPE \cite{metzler2000random}, and its solution can be compactly expressed by a particular  case of the Fox function \cite{metzler2000random}, also called the M-function \cite{mainardi2010m,pagnini2013m}:
\begin{equation}
P_B(x;t)= \frac{1}{\pi^{1/4}\sigma(t)} H_{1\,1}^{1\,0}\left[\frac{2|x|}{\pi^{1/4}\sigma(t)}\left|{\begin{array}{c}
      \left(3/4, 1/4\right)\\
      \left(0,1\right)\\
\end{array} } \right.\right].
\label{eq:FP_sol_x_Fox}
\end{equation}
Here, the MSD takes the form
\begin{equation}
    \sigma^2(t) = \frac{2D_x}{\sqrt{\pi D_y}}\sqrt{t}, 
\label{eq:Fox_MSD}
\end{equation}
but most importantly, in the limit $|x|/\sigma\gg 1$,   the Fox function in Eq.\eqref{eq:FP_sol_x_Fox} admits the asymptotic expansion \cite{mathai2009h,schneider1986stable}
\begin{equation}
 H_{1\,1}^{1\,0} \sim \left(\frac{2|x|}{\pi^{1/4}\sigma(t)}\right)^{-\frac{1}{3}}
e^{-\frac{3}{4^{4/3}\pi^{1/3}}\left(\frac{2|x|}{\sigma(t)}\right)^{\frac{4}{3}}}.
\label{eq:FP_sol_fisher}
\end{equation}
The expected Fisher's stretched-exponential behavior is indeed recovered, and 
remarkably, it reproduces the tails ($z\gtrsim 4$) of the numerical PDFs, reported in Fig.\ref{fig:pdfs_Fisher_tails}a, without any fitting procedure.

Extending this approach to the fractal combs of Fig.\ref{fig:cartoon} is out of the question, because the analogous of Eq.\eqref{eq:FP_COMB} cannot be drawn. 
However, we can pass through the CTRW representation. Let us first examine the comb model. If we consider only the dynamics along the backbone, the time spent at a site $x$ is the return time to $x$, after the walker has 
visited the matching SB. In other words, we can imagine the particle still being at site $x$, while performing its Brownian motion along the finger ($y$).
Within a certain degree of accuracy, we can assume that this residence (or waiting) time at $x$ corresponds to the walker's first return time to the backbone ($y=0$). Hence the residence time PDF $\omega(t)$ scales as $\omega(t) \sim (\tau/t)^{3/2}$, where $\tau$ is an arbitrary time-scale. 
On the other hand, the hopping dynamics along the backbone would ensure that 
the jump distribution is well approximated by a Gaussian, i.e., 
$\lambda(x)=(4\pi\sigma^2)^{-1/2}exp[-x^2/(4\sigma^2)]$. 
Following the derivation furnished in Ref.\cite{metzler2000random}, the 
Laplace and Fourier transforms of the waiting time and jump length PDFs read 
\begin{eqnarray}
  \omega(s)\sim 1-(s\tau)^{1/2}\,, \nonumber\\
  \lambda(k)\sim 1-(\sigma k)^2\,,
\end{eqnarray}
respectively.
Therefore the Fourier-Laplace transform of the CTRW PDF is
\[
P_B(k,s)=\frac{P_B(k,0)/s}{1+K_{1/2}s^{-1/2}k^2}
\]
so that the CTRW approach yields the correct MSD \eqref{eq:Fox_MSD} and, most importantly, it confirms that the corresponding Fokker-Planck equation is the 
FFPE \eqref{eq:FP_COMB} \cite{metzler2000random}.

Now let us turn to the fractal comb. The CTRW picture fully applies also to this case, with the only difference that a particle is imagined to reside at 
site $x$ while wandering through the fractal hinged on the backbone's site $x$ by its origin [see Fig.\ref{fig:cartoon}(A)]. 
Hence the residence (or waiting) time corresponds to the walker's first return time to the origin of a Sierpinski gasket of spectral dimension $d_s$, whose PDF behaves as \cite{gasketRW} 
$$
\omega(t) \sim (\tau/t)^{2-d_s/2}=(\tau/t)^{1+2\nu}\;.
$$
On the other side, the jump-length distribution is still a Gaussian. Thus, in this case, we have 
\begin{eqnarray}
  \omega(s)\sim 1-(s\tau)^{2\nu}\,, \nonumber\\
  \lambda(k)\sim 1-(\sigma k)^2 \,,
\end{eqnarray}
and the Fourier-Laplace transform of the walker's PDF is
$$
P_B(k,s)=\dfrac{P_B(k,0)/s}{1+K_{2\nu}s^{-2\nu}k^2}.
$$
Inverting, the proper Fokker-Planck equation for such a process reads 
\cite{metzler2000random}
\begin{equation}
\dfrac{\partial P_B(x;t)}{\partial t} = K_{2\nu}\dfrac{\partial^2 }{\partial x^2}{}_0D_t^{1-2\nu}P_B(x;t),
\label{eq:FP_sol_x_COMB}
\end{equation}
with the MSD 
\begin{equation}
\sigma^2(t) = \frac{2K_{2\nu}}{\Gamma(1+2\nu)}t^{2\nu}.
\label{eq:MSD_COMB_FR}
\end{equation}
As for Eq.\eqref{eq:FP_sol_x_total}, the solution of Eq.\eqref{eq:FP_sol_x_COMB} is again an M-function \cite{mainardi2010m,pagnini2013m} 
\begin{equation}
\begin{split}
P_B(x;t)= &
\frac{\sigma(t)}{\sqrt{2\Gamma(1+2\nu)}}\\  
& \times H_{1\,1}^{1\,0}\left[\sqrt{\frac{2}{\Gamma(1+2\nu)}}\frac{|x|}{\sigma(t)}\left|{\begin{array}{c}
      \left(1-\nu,\nu\right)\\
      \left(0,1\right)\\
\end{array} } \right.\right],
\label{eq:FP_sol_Fox_COMB}
\end{split}
\end{equation}
satisfying the asymptotic behavior  
\begin{equation}
\begin{split}
H_{1\,1}^{1\,0}\sim \left(\sqrt{\frac{2}{\Gamma(1+2\nu)}}\frac{|x|}{\sigma(t)}\right)^{\frac{2\nu-1}{2(1-\nu)}}\\times
e^{-(1-\nu)\nu^{\frac{\nu}{1-\nu}}\left(\sqrt{\frac{2}{\Gamma(1+2\nu)}}\frac{|x|}{\sigma(t)}\right)^{\frac{1}{1-\nu}}}\,,
\label{eq:FP_sol_fisher_COMB}
\end{split}
\end{equation}
which reproduces the Fisher's stretched exponential. 
In general, it can be shown that the M-function appearing in Eq.\eqref{eq:FP_sol_Fox_COMB} is related to the L\'evy stable distribution with parameter $\nu$ by the following equality \cite{pagnini2016stochastic}
\[
H_{1\,1}^{1\,0}\left[y\left|{\begin{array}{c}
      \left(1-\nu,\nu\right)\\
      \left(0,1\right)\\
\end{array} } \right.\right]=\frac{1}{\nu}L_\nu^{-\nu}\left(y^{-1/\nu}\right) y^{-(1+1/\nu)}.
\]
Figure \ref{fig:pdfs_Fisher_tails} shows that the asymptotic formula \eqref{eq:FP_sol_fisher_COMB} is an excellent approximation of the distribution tails.
Once again, we stress that no fit to numerical data has been implemented.

To sum up, expressions \eqref{eq:FP_sol_fisher}
and \eqref{eq:FP_sol_fisher_COMB} clearly demonstrate that Fisher's 
scenario is fulfilled if we take into consideration only the tails of the PDFs ($z\gtrsim 4$), and that it is a consequence of the type of fractional equation governing the diffusion on the comb backbone.

\section{Random-Shear Model \label{sec:shear}}
Random-shear models have been proposed to study the hydrodynamic transport of solute particles in stratified porous media, under the assumption that advection is parallel to the stratification planes \cite{mdem_model}.
Let $P(x,y,t)$ be the concentration of the solute particles and $\mathbf{U}=(U(y),0)$ be the shear parallel to $x$ depending only on the stratification height $y$, we can write the conventional advection-diffusion equation
\begin{equation}
    \dfrac{\partial P}{\partial t} + U(y) \dfrac{\partial P}{\partial x} =
D_x \dfrac{\partial^2 P}{\partial x^2} + 
D_y \dfrac{\partial^2 P}{\partial y^2}
\label{eq:RS_diff}
\end{equation}
where the incompressibility $\mathrm{div}(\mathbf{U})=0$ has been taken
into account and $D_{x},D_{y}$ denote the microscopic (molecular) diffusivity along $x,y$ directions respectively. In the following, we set $D_y=D_0$, because 
we will assume that $D_x\simeq 0$, so that the particle dispersion along $x$ 
is mainly ruled by the statistical properties of $U(y)$. 
In a famous paper, Matheron and de-Marsily \cite{mdem_model} studied the anomalous transport of particles driven by an array of horizontal random-width layers with a constant velocity. 
The particles also undergo a Brownian diffusion along the transverse direction, therefore, they cross different layers, visiting them several times. 
Specifically, Matheron and de-Marsily identified two properties of the velocity autocorrelation $\langle U(y) U(0) \rangle$ leading to a longitudinal anomalous super-diffusion, where the average is computed over the random field realizations.

The problem \eqref{eq:RS_diff} can be reformulated in terms of Langevin 
equations for an ensemble of independent particles
\begin{eqnarray}
    \dot{x}_i &=& U(y_i)                 \label{eq:shear_x}\\  
    \dot{y}_i &=& \sqrt{2D_0}\;\xi_i(t)  \label{eq:shear_y}  
\end{eqnarray}
$\{\xi_i(t)\}$ are independent, delta-correlated, and zero-mean Gaussian noises.
Periodic boundary conditions are enforced on the $y$-direction at $y=\pm L/2$, to implement a channel-like geometry along the $x$-axis.

Our shear longitudinal field is generated by a superposition of $M$ sinusoidal waves, i.e.
\begin{equation}
U(y) = \dfrac{1}{M}\sum_{k=\lambda}^{\Lambda} U_{k} \sin(k y + \phi_k) 
\label{eq:Ushear}
\end{equation}
where the sum on $k$ runs over the set $k=2\pi/L(1,2,\ldots,M)$,
with $\lambda=2\pi/L$ and $\Lambda=M\lambda$. 
This definition is not too restrictive for $M$ large, since any smooth-enough 
field can always be expanded in Fourier modes.
The amplitudes are assumed to depend on the  wave vectors as
\begin{equation}
U_{k} = U_0 |k|^{\gamma/2}, \quad k=\lambda,\ldots,\Lambda,
\label{eq:spectrum}    
\end{equation}
$U_0$ is a dimensional factor that can be set to unity by a simple time redefinition, while the presence of the phases $\{\phi_k\}$ is strongly 
necessary to confer a certain degree of heterogeneity (randomization) to the field. The parameter $\gamma$, defining the spectral properties of the modal decomposition of $U(y)$, is the only quantity that will be varied in our analysis. 
In what follows, it will be convenient to express $U(y)$ in the complex form
\begin{equation}
U(y) = \sum_{k=-\Lambda}^{\Lambda} {V_k} e^{i k y},
\end{equation}
with $V_{k}=U_{k} e^{i\phi_{k}}/(2iM)$, such that $V_{-k}=V^*_{k}$, 
$V_0=0$ and $\phi_{-k} = -\phi_k$. 
For reasons of convenience that will be soon clear, we take the shear field to be anti-symmetric about the middle of the channel, $y=0$, i.e. $U(-y)=-U(y)$. 
This condition also sets up the phase choice to $\phi_k = 0$ or $\pi$ with probability $(1/2,1/2)$, respectively.

The model (\ref{eq:shear_x},\ref{eq:shear_y}) can be exactly solved as the equation for $y$ is independent of $x$:  
the solution is $y(t) = y_0 + \sqrt{2D_0} w_t$, where $w_t$ indicates a Wiener's process, i.e. $\langle w_t \rangle = 0$ and $\langle w_s w_t \rangle = |t-s|$. 
A substitution into the first equation yields
\begin{equation}
x(t) = x(0) + \int_{0}^t\!ds\;U(y_0 + \sqrt{2D_0}\;w_s)
\label{eq:xsol}
\end{equation}
Thanks to Eq.\eqref{eq:Ushear}, Eq.\eqref{eq:xsol} expands to  
\begin{equation}
\begin{split}
x(t) - x(0) = & \\
              & \sum_{k=-\Lambda}^{\Lambda} V_{k} e^{i k y_0} \int_{0}^t\!ds\;\exp[ik\sqrt{2D_0}\;w_s].
\label{eq:x_of_t}
\end{split}
\end{equation}
Now, from expression \eqref{eq:x_of_t}, all the moments 
$$
M_m(t) = \langle [x(t) - x(0)]^m \rangle
$$ 
can be computed. 
The crucial point is to establish the meaning of the average $\langle\cdots\rangle$. 
As a matter of fact, three types of average can be carried out on this system:
\begin{enumerate}
    \item over the noise realizations, i.e. on the Wiener's process $w_t$: $\langle\ldots\rangle_w$
    \item over the initial conditions $y_0$ along the stripe: $\langle\ldots\rangle_0=\int_{-L/2}^{L/2} dy_0 \, \rho(y_0)$
    \item over the possible configurations of the disordered field $U(y)$, i.e. over the phases: $\langle\ldots\rangle_\phi= \prod_{k=\lambda}^\lambda\int_{-\pi}^{\pi} d\phi_k/(2\pi)$.
\end{enumerate}
 Unlike previous works on the random-shear model \cite{dreizin1973anomalous,Bouchaud,Ben-Avraham,majumdar2003,mdem_model,bakunin2008turbulence, bakunin2011chaotic, compte1998fractional,rednerPRL, gaveau1992anomalous, gaveau1988anomalous,mazo1998taylor,roy2006motion,zumofen1990enhanced}, in our analysis we will not consider the average over random field realizations (phases in our case), 
 taking only the average on the Wiener process and on the initial conditions, 
 as done in Ref.\cite{dentz2008}, in symbols $\langle\langle\cdots\rangle_w\rangle_0$. We will be assuming an initial uniform distribution of particles along the channel: $\rho(y_0)=1/L$.

First we can compute the mean displacement (drift),
\begin{equation}
\begin{split}
\langle\langle[x(t)-x(0)]\rangle_w\rangle_0 =& \\
& \sum_{k=-\Lambda}^{\Lambda} \dfrac{V_k}{D_0 k^2} \langle e^{i k y_0}\rangle_0 \bigg(1 - e^{-D_0 k^2 t}\bigg)
\end{split}
\label{eq:drift0}
\end{equation}
which 
is rigorously zero because $\langle\exp(i k y_0)\rangle_0=\delta_{k,0}$ and we have used the property of $w_t$ such that $\langle\exp(i k w_t)\rangle_w = \exp(-D_0 k^2 t)$.   

The mean square displacement reads
\begin{equation}
\begin{split}
\langle\langle[x(t)-x(0)]^2\rangle_w\rangle_0 = & 
 \bigg(\sum_{k=\lambda}^{\Lambda} \dfrac{|V_k|^2}{D_0 k^2}\bigg)\;t \\
& -\sum_{k=\lambda}^{\Lambda} \dfrac{|V_k|^2}{D_0^2 k^4} \bigg(1 - e^{-D_0 k^2 t}\bigg)\,; 
\label{eq:MSD}
\end{split}
\end{equation}
recalling that $V_k = U_k e^{\phi_k}/(2iM)$, we have $|V_k|^2 = U^2_k/(4M^2)$, so that the MSD turns out to be independent of the phase disorder. 
The first contribution is the well-known Taylor term \cite{taylor1953dispersion}
\begin{equation}
D_{\mathrm{eff}} = \dfrac{1}{2}\sum_{k=\lambda}^{\Lambda} \dfrac{|V_k|^2}{D_0 k^2}
\label{eq:taylor}
\end{equation}
of the effective standard diffusion, whereas the second term 
is the contribution leading to the anomalous transient behavior whose duration
increases with the width (transversal size) $L$ of the channel. 

Before showing analytically how the anomalous regime emerges from 
Eq.\eqref{eq:MSD},
we repeat the matching argument of sec.\ref{sec:comb}, in the context of the random-shear model to obtain its anomalous exponent in an intuitive way.  
Let us consider the case of large but finite $L$; of course, 
when $t\gtrsim t_*=L^2/D_0$, the transverse free diffusion of particles feels the boundary effects, and the longitudinal diffusion changes from anomalous to standard.
Therefore, even for the random-shear, we can assume that Eq.\eqref{eq:arg2} 
holds true, with the exponents determined by the random shear problem. 
In this case, $D(L)$ scales as   
$$
D_{\mathrm{eff}}(L) \sim \dfrac{L^{1-\gamma}}{D_0},    
$$
according to the Taylor formula \eqref{eq:taylor}, where the definition, Eq.\eqref{eq:spectrum}, has been used along with an implicit passage from the summation on $k$ to the integral over $dk$.

The anomalous and the standard regimes need to match each other at time $t_*(L)$, which is the typical time, $t_*(L)\sim L^2/D_0$, taken by transverse free 
diffusion [Eq.\eqref{eq:shear_y}] to become almost uniform over the channel width $L$.
Thus, from the matching condition, $t_*(L)^{2\nu} \sim D_{\mathrm{eff}}(L) t_*(L)$, one obtains the relation $(L^2)^{2\nu} \sim L^{1-\gamma} L^2$, which, by equating the exponents, leads to
\begin{equation}
\nu = \dfrac{3-\gamma}{4}.
\label{eq:nu_anomal}
\end{equation}
Therefore, as in the case of comb structures, a simple scaling argument links the anomalous exponent $\nu$ to the parameter $\gamma$  defining the velocity field spectrum through Eq.\eqref{eq:spectrum}, in addition Eq.\eqref{eq:fisheralpha} provides the value 
\begin{equation}
\alpha = \frac{4}{1+\gamma}
\label{eq:fishershear}
\end{equation}
for the Fisher's parameter.

The above derivation can be made rigorous, again assuming large channel widths 
$L\gg 1$, so the summation over $k$ can be replaced by an integral over the interval $[\lambda,\Lambda]$, casting Eq.\eqref{eq:MSD} in the following form:
\begin{equation}
\begin{split}
\langle\langle[x(t)-x(0)]^2\rangle_w\rangle_0  = & 
 2D_{\mathrm{eff}} t  \\
& -\frac{L}{2\pi}\frac{U_0^2}{D_0^2} \int_{\lambda}^{\Lambda}\!\!dk\;k^{\gamma-4} (1 - e^{-D_0 k^2 t}),
\label{eq:intMSD}
\end{split}
\end{equation}
where Eq.\eqref{eq:spectrum} has been used. 
Appendix \ref{app:A} shows that, in the time interval $L^2/(M^2 D_0)\ll t\ll L^2/D_0$, the mean-square displacement is
\begin{equation}
\langle\langle[x(t)-x(0)]^2\rangle_w\rangle_0 =
 2D_{\mathrm{eff}} t  + B\,(D_0 t)^{\frac{3-\gamma}{2}} - A\,,
\label{eq:anoMSD}
\end{equation}
where the constants are:
$A =\Lambda^{\gamma-3}/[\lambda(\gamma-3)] \,(U_0/D_0)^2$, and 
$B=2 \Gamma[(\gamma+1)/2]/[\lambda(\gamma-3)(\gamma-1)]\,(U_0/D_0)^2$.
Equation \eqref{eq:anoMSD} prescribes that a particle driven by the random shear
undergoes a transient super-diffusion with the exponent \eqref{eq:nu_anomal}.  
Of course, for $t\gg L^2/D_0$ the MSD turns into the Gaussian regime characterized by a coefficient \eqref{eq:taylor}.

\subsection{Numerical results}
We numerically simulated the motion of $N=2\times 10^6$ particles evolving 
according to Eqs.(\ref{eq:shear_x},\eqref{eq:shear_y}) with $D_0 = 0.1$.
The system is prepared into an initial distribution of $\{x_i(0),y_i(0)\}$, 
$i=1,\ldots,N$
that is equally spaced along the channel width, $y_{i}(0)=-L/2 + L(i-1)/(N-1)$,  
and uniformly distributed in $x$ on the interval $[-0.5,0.5]$.
The equal spacing along $y$ guarantees that the constraint 
$\langle e^{i k y_0}\rangle_0 = \delta_{k,0}$ is numerically well satisfied by the initial condition.

The shear parameters are kept fixed to: $U_0=1$, $L=100$, and 
$M=100$, while $\gamma$ in Eq.\eqref{eq:spectrum} is varied in the range $-1 \le\gamma\le 1$.
Moreover, our specific choice of the disorder ($\phi_k=0$ or $\pi$) 
corresponds to assign random signs to the amplitudes $U_k = \pm U_0|k|^{\gamma/2}$.
The numerical integration of Eqs.(\ref{eq:shear_x},\ref{eq:shear_y}) has been performed with a simple Euler scheme with a time step $h = 0.01$. 

\begin{figure}[ht]
\centering
\includegraphics[width=0.85\columnwidth]{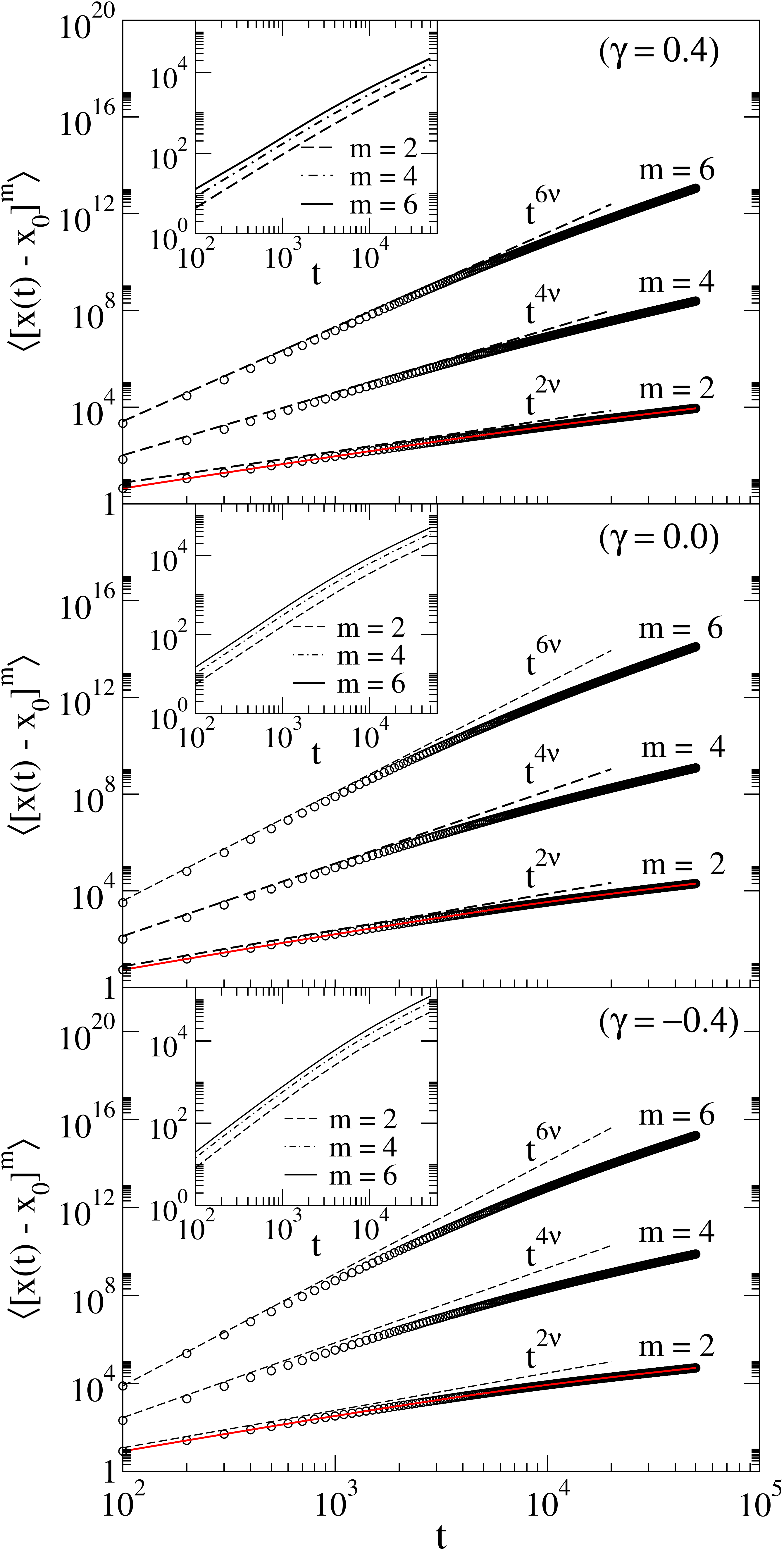}
\caption{
Time behavior of the moments $M_{2m}(t) = \langle\langle [x(t) - x(0)]^{2m} \rangle_w\rangle_0$ ($m=1,2,3$) of $N=2\times 10^6$ particles in the shear model 
with parameters, $L=100$, $M=100$, $D_0 = 0.1$, $U_0=1$, and $\gamma=(-0.4, 0.0, 0.4)$. 
The red solid curves indicate Eq.\eqref{eq:MSD}, and the dashed 
straight lines are the expected anomalous scaling, $2\nu m$, with $\nu=(3-\gamma)/4$.   
The insets show the alignment of the moments upon raising  
$[M_{2m}(t)]^{1/m}$, which is a strong numerical indication of the 
collapse of the particle distributions, Eq.\eqref{eq:PDFscale}. 
\label{fig:mom}
}
\end{figure}

The first test of the correctness of the Euler integration is the agreement of the MSD obtained by the simulations with the analytical result, Eq.\eqref{eq:MSD}, as verified in Fig.\ref{fig:mom} 
by the coincidence of the red curves representing Eq.\eqref{eq:MSD} with the numerical MSD (open circles).
We computed also the time behavior of the moments $M_m(t)$ with $m=4,6$ to verify their expected intermediate anomalous behavior \eqref{eq:Mom_vs_t} occurring in the interval $L^2/(M^2 D_0)\ll t\ll L^2/D_0$, which is represented by the  dashed lines in Fig.\ref{fig:mom}. 
For times large enough, $t \gtrsim 2\times 10^4$, the Gaussian scaling of moments is recovered, with the normal exponent $\nu=1/2$, indicating that the system has attained the standard regime.
 
The insets in Fig.\ref{fig:mom} show also the alignment of moments  
upon raising them to the right power, 
$[M_{2m}(t)]^{1/m} \sim M_2(t)$ (Fig.\ref{fig:mom}). 
This alignment is a consequence of the scaling \eqref{eq:PDFscale}, 
and the data in the insets represent a first robust numerical test that this scaling is verified by the simulations. 
The same conclusion was reached in Refs.\cite{Bouchaud,Ben-Avraham} 
using the average over the disordered convection fields 
$\langle\cdots\rangle_\phi$.  

\begin{figure}[ht]
\centering
\includegraphics[width=0.85\columnwidth]{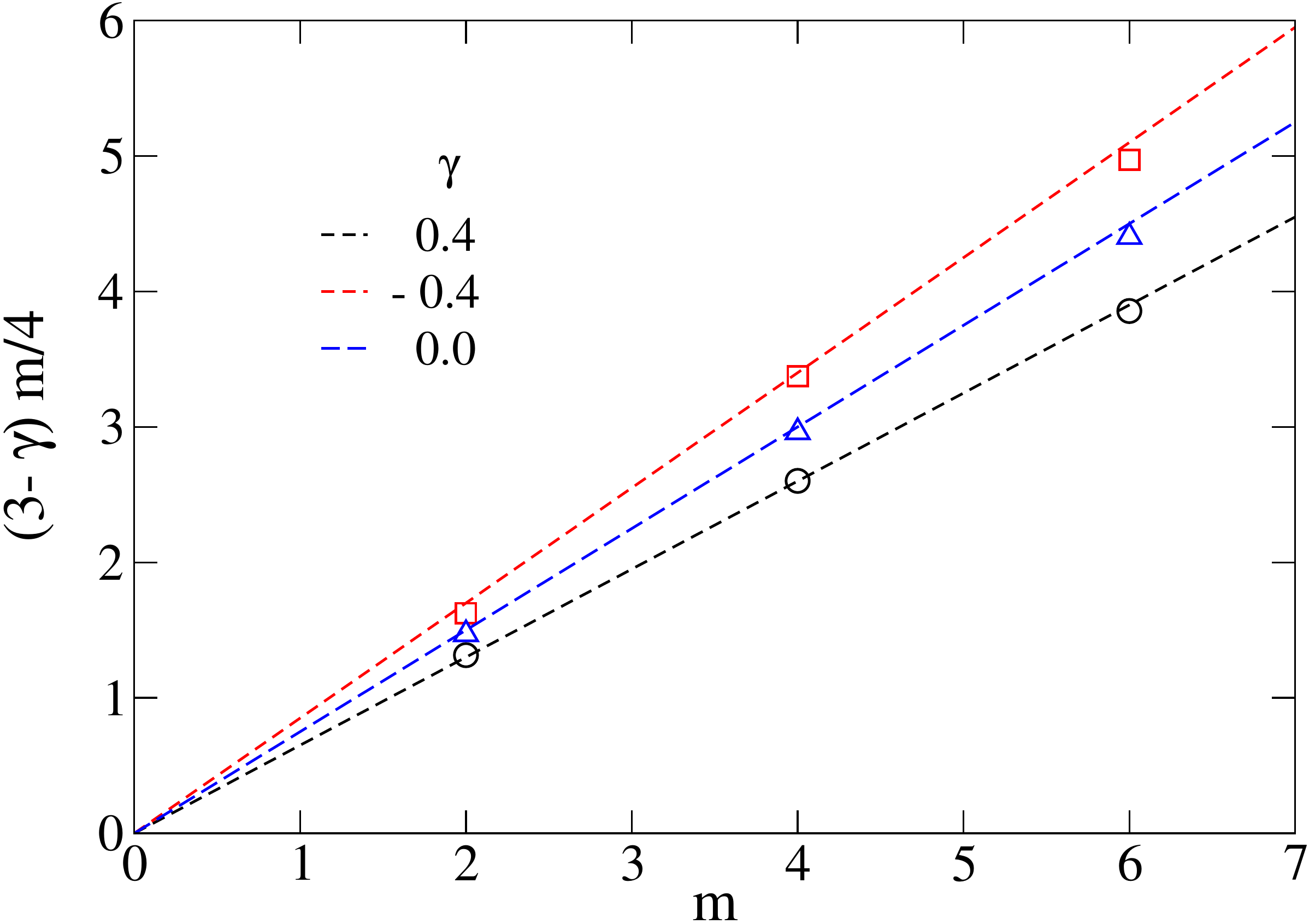}
\caption{Plot of the anomalous spectrum of moments,
$\alpha(m)$, vs the order of the moments $m$.
The simple linear behavior indicates that diffusion is not 
strongly anomalous.
\label{fig:spectrum}
}
\end{figure}

The collapse of moments in Fig.\ref{fig:mom} shows the not strongly 
anomalous character of the superdiffusive regime \cite{Castiglione99}. 
In view of Eq.\eqref{eq:strong_anomal} indeed, we have $\nu(m)\equiv \nu =(3-\gamma)/4$, as is shown in Fig.\ref{fig:spectrum} for the same values of $\gamma$ displayed in Fig.\ref{fig:mom}.
Now, as the scaling of the moments corroborates the validity of Eq.\eqref{eq:PDFscale}, we need to establish whether the 
tails of the PDFs satisfy Eq.\eqref{eq:Stretched_exp} with the expected 
exponent $\alpha = 4/(1+\gamma)$. 

\begin{figure}[ht!]
\centering
\includegraphics[width=0.85\columnwidth]{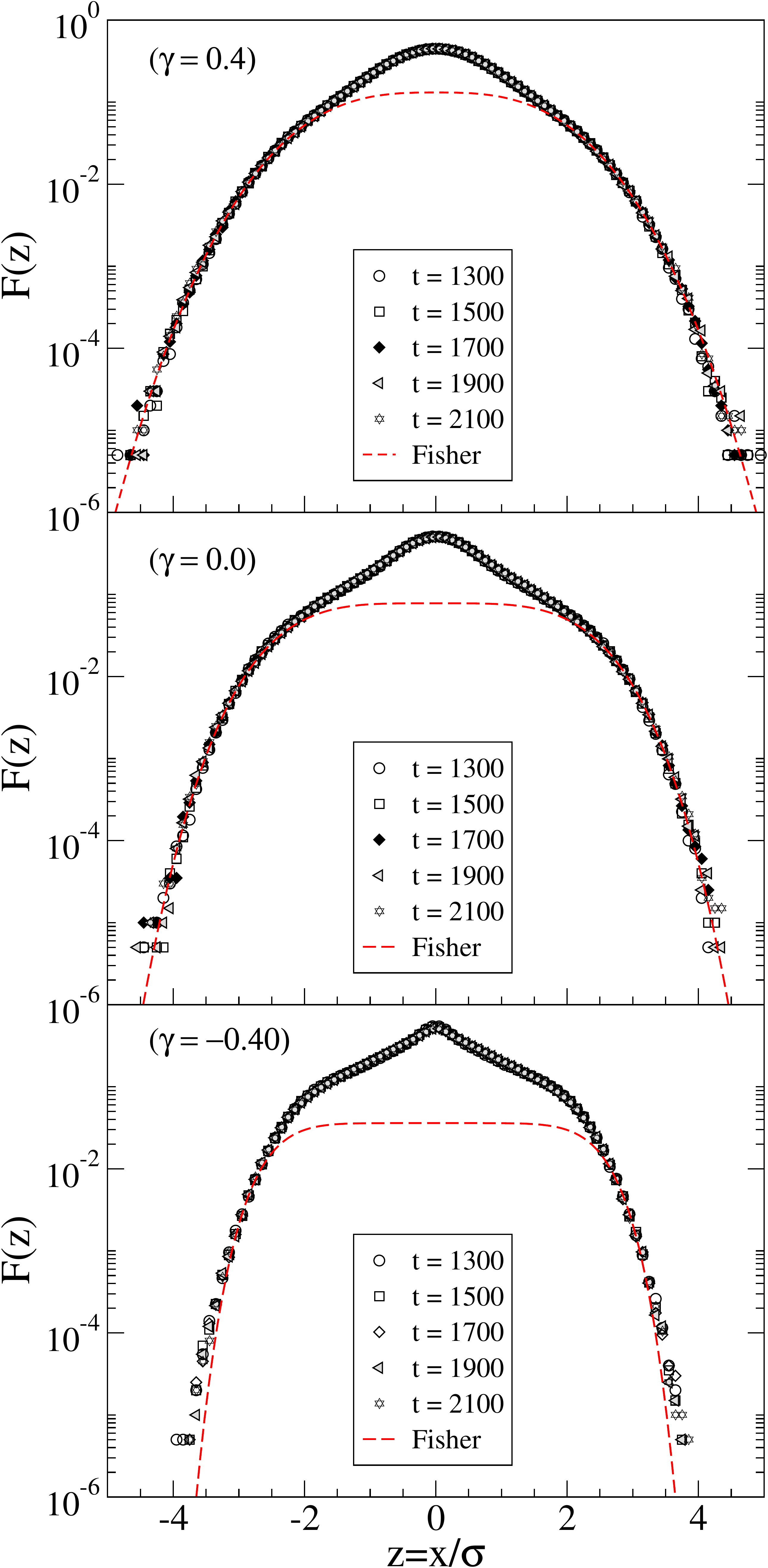}
\caption{Collapse of the PDF, Eq.\eqref{eq:PDFscale}, upon the rescaling $x\to x/t^{\nu}$, for $\gamma={-0.4,0.0,0.4}$. 
The dashed red curves represent the fit of the tails with the Fisher's scaling function, with $\alpha = 4/(1+\gamma)$.
Notice that the case $\gamma=0$, for which $\nu=3/4$, corresponds to 
the Matheron de-Marsily model \cite{mdem_model}.
The last panel, $\gamma=-0.4$, suggests that the Fisher's fitting is not satisfactory for negative $\gamma$.
\label{fig:scaling}
}
\end{figure}

Figure \ref{fig:scaling} displays the collapse of the PDF at different times according to Eq.\eqref{eq:PDFscale}.
Moreover, the red dashed curves represent the fitting with Eq.\eqref{eq:Stretched_exp}, with the constraint $\alpha=4/(1+\gamma)$ being priorly 
imposed.
In practice, only the amplitude, $A$, and the parameter, $a$, are adjusted by the fitting procedure.
Although the fitting curve fails to reproduce the bulk of the simulated PDF, it is very reasonable for the far tails. 
Besides the values $\gamma=0, 0.4$, the consistency of the Fisher's scenario has been also verified for $\gamma=0.2,0.6,0.8$ (not shown).

Notice that the case $\gamma=0$, implying $\nu=3/4$, corresponds to 
the Matheron de-Marsily model \cite{mdem_model}. 

However, the Fisher's prediction seems to fail for the PDF with negative 
$\gamma$; see the red dashed curve in the bottom panel of Fig.\ref{fig:scaling}.Although the tails of the rescaled distributions are still stretched 
exponential, the exponent is different from $\alpha=4/(1+\gamma)$.

\begin{figure}[ht!]
\centering
\includegraphics[width=0.85\columnwidth]{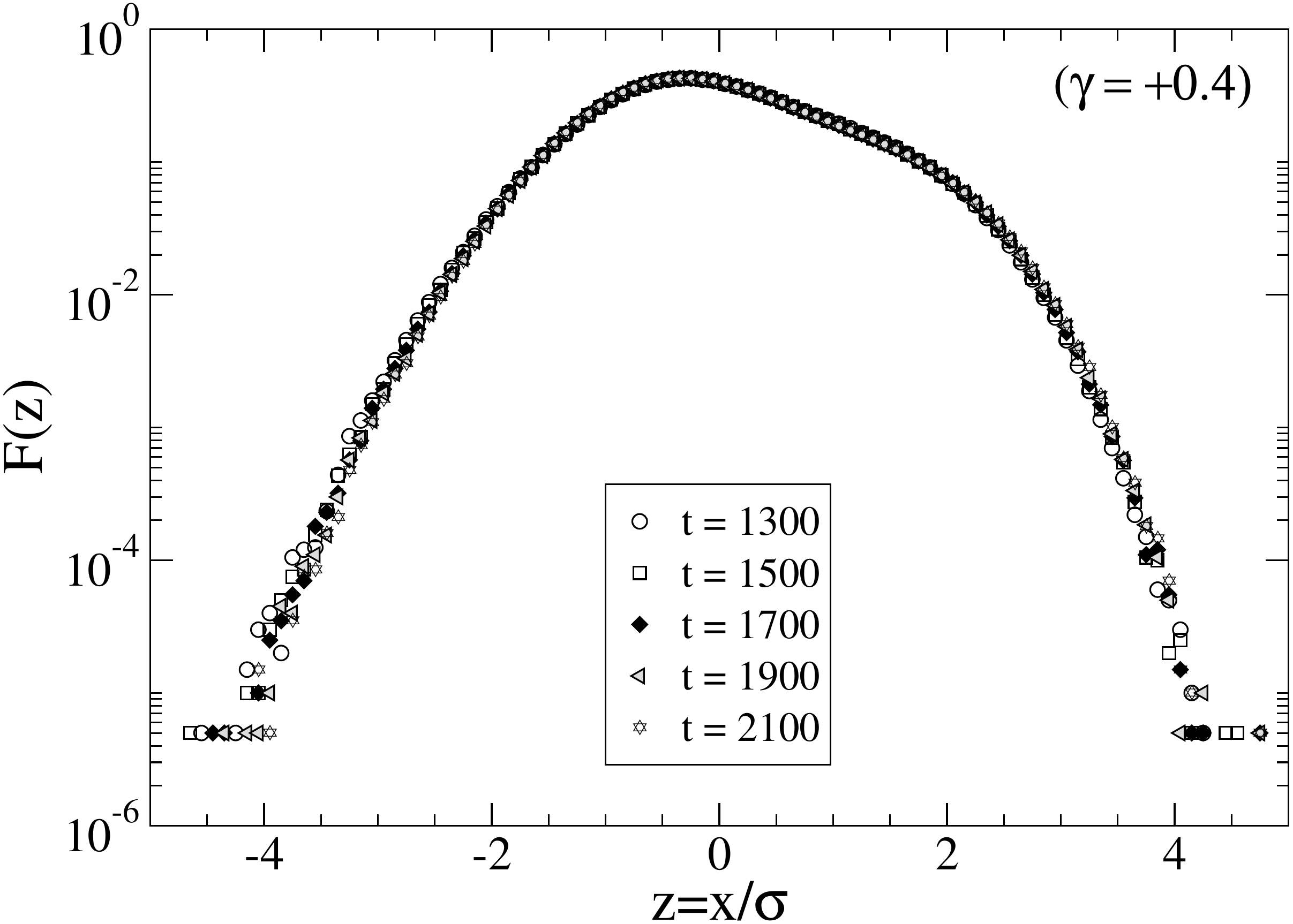}
\caption{
Collapse of the PDF along $x$ of the particles subjected to a
shear field with a phase disorder $\phi = \pm \pi/2$ and $\gamma=0.4$.
The collapse of the PDF at different times indicates that the scaling, 
Eq.\eqref{eq:PDFscale}, is verified, however, the PDF, both for symmetry and 
tails, is not in the Fisher's class.
\label{fig:pdfcos} 
}
\end{figure}

As a final remark, we can say that the behavior of the longitudinal 
PDF is strongly dependent on the properties of the shear field, so 
the Fisher scenario is not so robust. For instance, if the
phases are such that $\phi = \pm \pi/2$, with probability $1/2$, 
we obtain a distribution that still follows the scaling law
\eqref{eq:PDFscale}. However the tails are not Fisher-like 
(see Fig.\ref{fig:pdfcos}) also because a generic disorder does not 
preserve the symmetry, $x\to -x$, of the Fisher's distributions.

\section{Conclusions
\label{sec:concl}}
In this work we have studied the subdiffusion along the backbone 
of random walks on comb-like fractal structures and the superdiffusion 
of the Lagrangian dynamics of Brownian particles in a random-shear velocity 
field. 
In this case the anomalous transport occurs along the shear direction.

In the comb systems, the anomalous exponent can be analytically 
obtained as a function of the fractal dimension of the Sierpinski 
sidebranches decorating the backbone. 
 In the shear model the anomalous exponent is a function of the parameter 
$\gamma$ defining the spectral properties of the Fourier-mode combination of the velocity field. 

We focused our analysis mainly on the scaling properties of the spatial PDF of the process along the preferential transport direction: 
the backbone (for combs) and $x$-axis (for random shear), because we were 
interested in testing the assumption that such PDFs develop 
Fisher's stretched-exponential tails $\exp(-a|z|^{\alpha})$, with $\alpha$ related to the anomalous exponent by the relation $\alpha=1/(1-\nu)$.

Our simulations show that for comb systems, the PDF of the anomalous sub-diffusion along the backbone follows the Fisher's tails, regardless of the complexity of the fractal sidebranches. This numerical finding has been also supported by analytical predictions based on the extension 
of the fractional Fokker-Planck equation to comb fractal structures.

For the random shear, we found that the transport is anomalous with an exponent $\nu=(3-\gamma)/4$. The particle PDFs along the $x$-axis exhibit 
stretched-exponential tails with an exponent consistent with the Fisher's prediction, Eq.\eqref{eq:fishershear}, only for $\gamma>0$. Surprisingly, for $\gamma<0$, numerical simulations seem to indicate that the Fisher's scenario breaks down.

\section*{Acknowledgement}
F.C. and A.V. acknowledge the financial suppor from MIUR, PRIN project 
n.201798CZL ``Coarse-grained description for non-equilibrium systems and 
transport phenomena (CO-NEST)''.

\newpage
\bibliographystyle{unsrt}
\bibliography{references}

\appendix
\section{Analytical derivation of the  MSD \label{app:A}}
In this appendix, we derive Eq.\eqref{eq:anoMSD} by solving the integral 
\begin{equation}
I=\frac{L}{2\pi}\frac{U_0^2}{D_0^2} \int_{\lambda}^{\Lambda}\!\!dk\;k^{\gamma-4} 
(1 - e^{-D_0 k^2 t})
\end{equation}
appearing in Eq.\eqref{eq:intMSD}.
A first integration by parts yields

\begin{equation}
\begin{split}
I=\frac{L}{2\pi(\gamma-3)}\frac{U_0^2}{D_0^2}\;
\bigg\{ 
\bigg[k^{\gamma-3}(1-e^{-D_0 k^2t})\bigg]_{\lambda}^{\Lambda} \\
-2D_0t\int_{\lambda}^{\Lambda}\!\!dk\;k^{\gamma-3} e^{-D_0 k^2 t}
\bigg\}
\end{split}
\end{equation}

Then we apply the change of variable $y=D_0k^2t$ in the remaining integral, achieving the final expression
\begin{equation}
\begin{split}
I&=\frac{L}{2\pi(\gamma-3)}\frac{U_0^2}{D_0^2}\times\\
&\left\{\Lambda^{\gamma-3}\left(1-e^{-\Lambda^2D_0t}\right)-\lambda^{\gamma-3}\left(1-e^{-\lambda^2D_0t}\right)\right.\\
&\left.
-(D_0t)^{\frac{3-\gamma}{2}}\left[\Gamma\left(\frac{\gamma-1}{2},\lambda^2D_0t\right)-\Gamma\left(\frac{\gamma-1}{2},\Lambda^2D_0t\right)\right]\right\},
\end{split}
\end{equation}
where $\Gamma$ is the  incomplete gamma function \cite{abramowitz1988handbook}. For intermediate times such that $\frac{1}{\Lambda^2D_0}\ll t\ll \frac{1}{\lambda^2D_0}$, we recall that $\Gamma(a,x)\sim x^{a-1}e^{-x }$ as $x\to \infty$ and $\Gamma(a,x)\sim\Gamma(a)-\frac{x^a e^{-x}}{a}$  for $x\to 0$ \cite{abramowitz1988handbook}, hence the expression reported in Eq.\eqref{eq:anoMSD} is recovered after straightforward algebraic manipulations.


\end{document}